\documentclass{blois}

\usepackage{textcomp}

\def\Journal#1#2#3#4{{#1} {\bf #2}, #3 (#4)}

\def\PRL{\em Phys. Rev. Lett.}
\def\PRD{{\em Phys. Rev.} D}

\def\be{\begin{equation}}
\def\ee{\end{equation}}
\def\bea{\begin{eqnarray}}
\def\eea{\end{eqnarray}}

\newcommand{\Photo}{}

\begin{document}
\begin{flushright}
DESY 17-208
\end{flushright}

\vspace*{4cm}
\title{GENERATION OF ASYMMETRIC DARK MATTER \\ AND GRAVITATIONAL WAVES \footnote{Talk based on {\em JCAP} {\bf 1705}, 28 (2017)}}

\author{ I. BALDES }

\address{ DESY, Notkestra{\ss}e 85, D-22607 Hamburg, Germany}

\maketitle\abstracts{We consider the possibility of a gravitational wave signal in an asymmetric dark matter model. In this model a generative sector produces both the baryon asymmetry and a dark matter asymmetry in a strong first-order phase transtion. Bubble collisions during the phase transition lead to sound waves in the plasma which are a source of a stochastic gravitational wave background. We consider the prospects of future graviational wave observatories such as LISA and BBO detecting such a signal. Constraints on the model from Halo ellipticity, $\Delta N_{\rm eff}$ and direct detection experiments are also discussed.}

\section{Introduction to Asymmetric Dark Matter}
To motivate the consideration of asymmetric dark matter, it is first instructive to consider the visible sector. In the visible sector the matter density is set by a particle-antiparticle asymmetry, i.e. the baryon asymmetry. The working hypothesis is that a dynamical process of baryogenesis created the baryon asymmetry, $n_{B} \equiv n_{b}-n_{\overline{b}}$, where $n_{b}$ is the baryon density and $n_{\overline{b}}$ is the antibaryon density. In setting the visible sector matter density it is also important that the baryons and antibaryons annihilate efficiently. After confinement of QCD this occurs through nucleon-antinucleon annihilation. Indeed because this annihilation is so efficient the relic antibaryon density is completely negligible and the relic baryon density is set by the asymmetry: $n_{b} \simeq n_{B}$. The visible sector matter density can therefore be written as $\rho_{\rm vis} =(n_{b}+n_{\overline{b}})m_{p} \simeq n_{b}m_{p} \simeq n_{B}m_{p}$ where $m_{p}$ is the proton mass.

It is logical to consider models in which the dark sector density is set in a similar way: these are the so called asymmetric dark matter models.~\cite{ADMreview} This necessitates the creation of an asymmetry in the dark sector, $n_{D} \equiv n_{\rm dm} - n_{\overline{\rm dm}}$, completely in analogy with baryogenesis, together with efficient annihilation of the symmetric dark matter component, in analogy with the process of nucleon-antinucleon annihilation. The dark matter density can then be written as $\rho_{dm} =(n_{\rm dm}+n_{\overline{\rm dm}})m_{\rm dm} \simeq n_{\rm dm}m_{\rm dm} \simeq n_{D}m_{\rm dm}$, where $m_{\rm dm}$ is the DM mass. Of course one is then free to speculate that the dark and visible sector asymmetries are somehow related, either created together in a generative sector, as is the case for the model discussed below, or through a transfer mechanism between the visible and dark sectors.

\section{Asymmetric Dark Matter from a Generative Sector}
\subsection{The model}
\begin{figure}[t]
\begin{center}
\includegraphics[width=130pt]{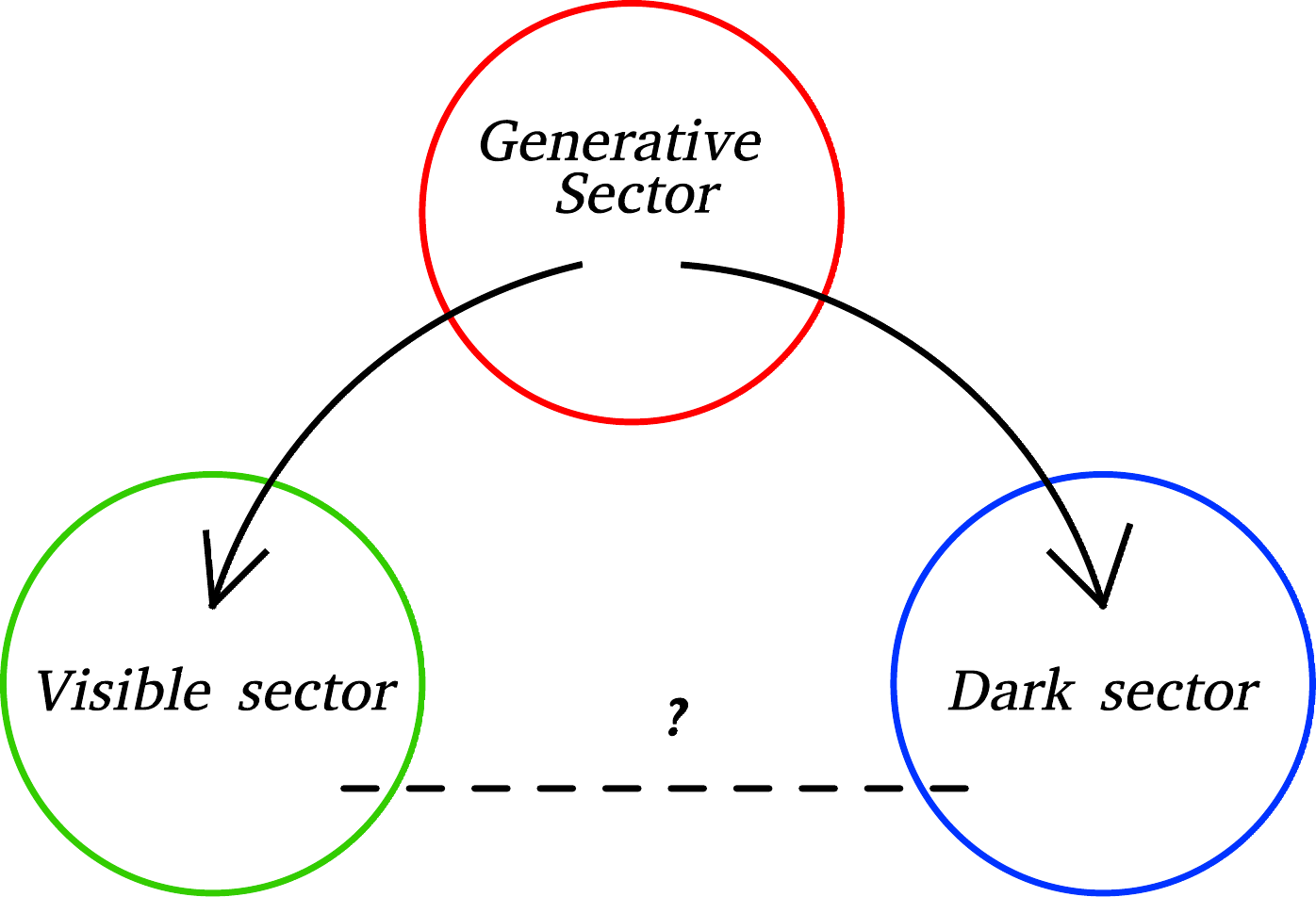}
\end{center}
\caption{\small Diagramatic depiction of the model structure.}
\label{fig:sectors}
\end{figure}

Currently from particle physics we know the SM in detail, which consists of the gauge group $SU(3) \times SU(2) \times U(1)$ together with the fermionic matter content and the Higgs. This is a non-minimal structure with chiral fermions and global $B+L$ anomaly. It contains the necessary ingredients for electroweak barogenesis. In such a scenario the Higgs field undergoes a strong first-order phase transition. Charge Parity-violating collisions of fermions with the bubble walls lead to a chiral asymmetry forming on the outside of the bubbles. Sphalerons would then convert this to a baryon asymmetry. This is swept into the expanding bubble where sphalerons are suppressed. However, the phase transition is the SM is a crossover and there is also insufficient CP violation to explain the observed $n_{B}$. However we may ask ourselves if similar beyond-the-SM physics could exist? A first-order phase transition in a generative sector could then produce the baryon and a DM asymmetry. Such a phase transition will also result in gravitational waves. The Petraki, Trodden and Volkas scenario~\cite{Kallia} is a concrete realisation of such a picture and the focus of this talk. The overall structure of the model consists of three sectors and is depicted in Fig.~\ref{fig:sectors}. Let us quickly summarise the model.

The generative sector consists of an SU(2)$_{G}$ gauge group, a scalar $\phi \sim 2$ which breaks the gauge symmetry in a strong first-order phase transition and chiral fermions $\Psi_{L} \sim 2$ and $\Psi_{R} \sim 1$.
The sector has a global anomaly which, together with the Yukawa interactions,
	\begin{equation}
	\mathcal{L} \supset -\frac{1}{\sqrt{2}}\left(\sum_{j=1}^{2}h_{j}\overline{\Psi_{L}}\phi\Psi_{jR}+\tilde{h}_{j}\overline{\Psi_{L}}\tilde{\phi}\Psi_{jR} \right)+H.c.,
	\label{eq:genyuk}
	\end{equation}
which act as a source of CP violation can lead to a particle-antiparticle asymmetry forming in complete analogy with electorweak baryogenesis. The asymmetry is transfered to the other sectors via a sequence of interactions starting with
	\begin{equation}
	\mathcal{L} \supset -\frac{\kappa}{\sqrt{2}}\overline{\Psi_{L}}\chi f_{R} +H.c. 
	\end{equation}
where $\chi \sim 2$ under SU(2)$_{G}$ is a scalar and $f_{R}$ the right-chiral component of a sterile Dirac fermion. The asymmetry is communicated to the visible sector via
	\begin{equation}
	\mathcal{L} \supset -\frac{y_{1}}{\sqrt{2}}\overline{l_{L}}Hf_{R} + H.c.,
	\end{equation}
where $l_{L}$ is the SM lepton doublet and $H$ is the SM Higgs, and to the dark sector via
	\begin{equation}
	\mathcal{L} \supset -\frac{y_{2}}{\sqrt{2}}\overline{\xi}\chi\zeta + H.c..
	\end{equation}
Here DM consist of the Dirac fermions $\zeta$ and $\xi$. As the dark sector asymmetry is equal to the visible sector asymmetry times a sphaleron reprocessing factor we are left with the mass relation $m_{\zeta}+m_{\xi}\approx 1.5$ GeV. The symmetric component of the DM is annihilated away with a dark $U(1)_D$, under which $\zeta$ and $\xi$ are charged.

\subsection{Gravitational Waves}
\begin{figure}[t]
	\begin{center}
	\includegraphics[width=180pt]{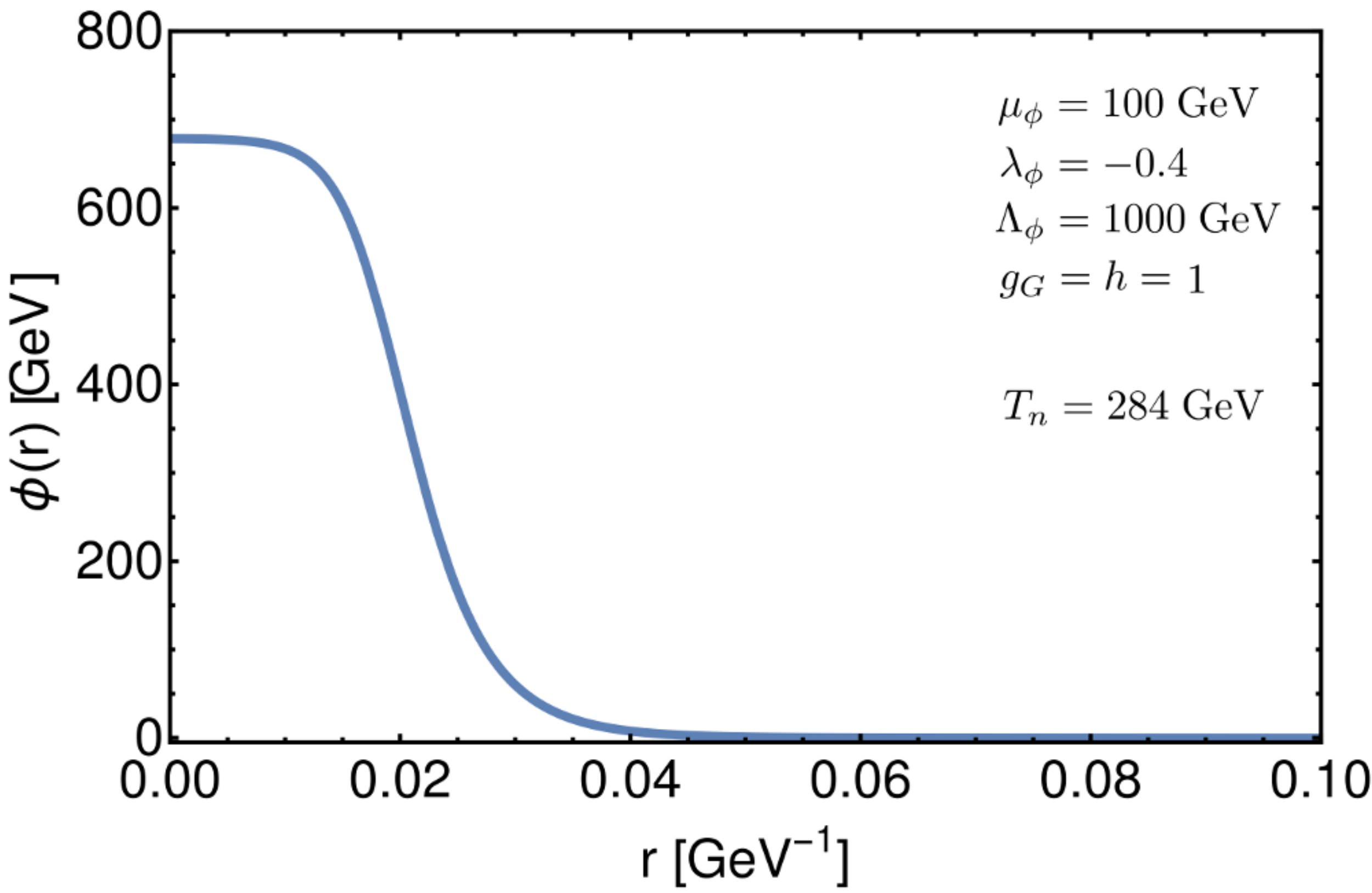} $\qquad$
	\includegraphics[width=180pt]{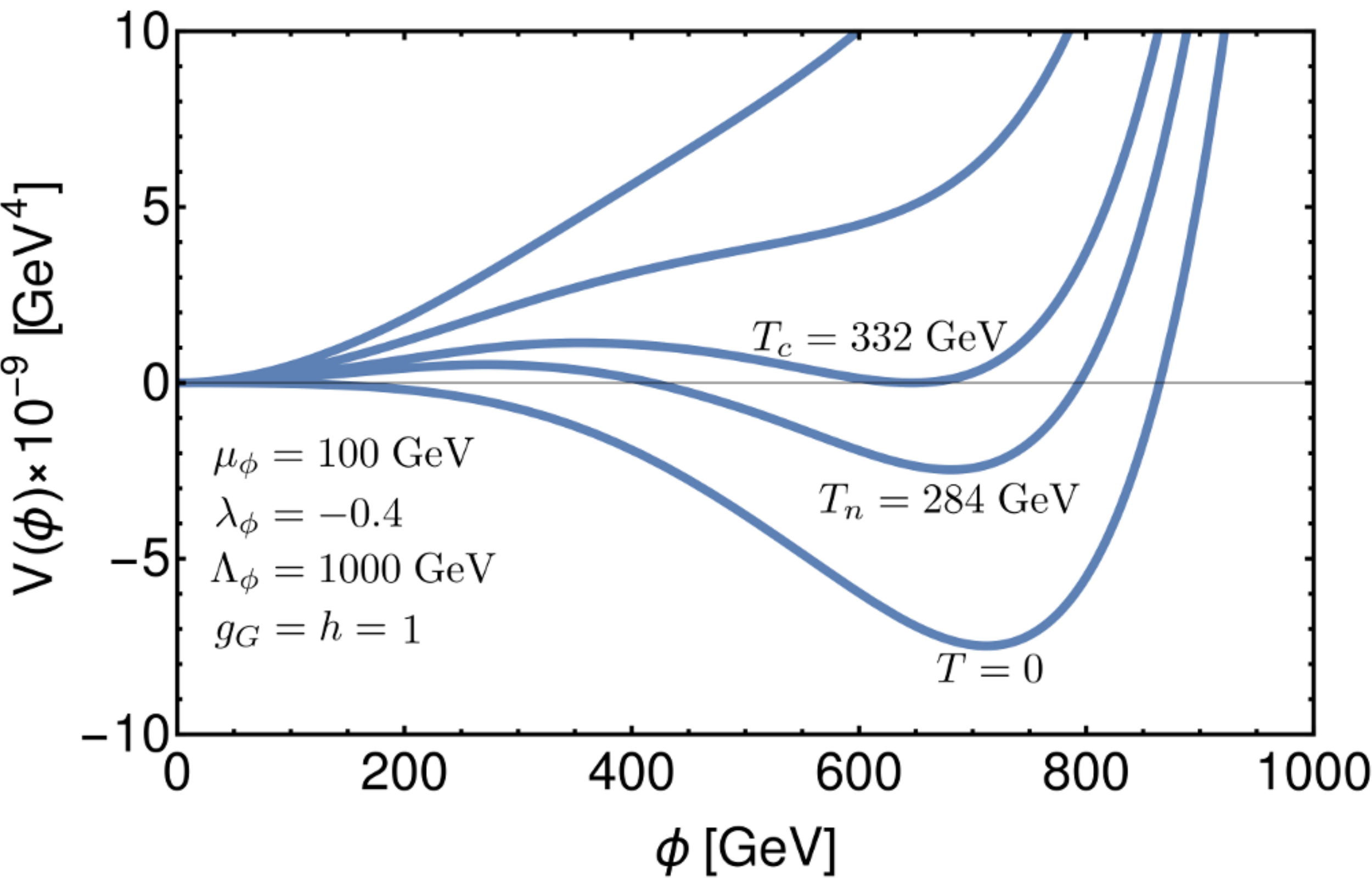}
	\end{center}
\caption{\small Left: example of the bubble profile at nucleation. Right: illustration of the temperature evolution of the potential. Nucleation may occur once the minimum at finite field values drops below the symmetric phase minimum at the origin.}
\label{fig:bubble}
\end{figure}

Once the field content of the model is specified, it is clear that a first-order phase transition can be arranged for $\phi$, allowing for the generation of an asymmetry. Therefore the details of the phase transition was not studied in the original paper.~\cite{Kallia} In constrast, here we are certainly interested in the details of the phase transition as this will allow us to calculate the expected gravitational wave signal. We must therefore specify a potential. Here we consider a simple possibility with
	\begin{equation}
	V_{G} = \frac{ \mu_{\phi}^{2} }{ 2 }\phi^{2} + \frac{ \lambda_{ \phi }}{4} \phi^{4} + \frac{1}{8\Lambda_{\phi}^{2}}\phi^{6}.
	\label{eq:v6pot}
	\end{equation} 
The strong phase transition is achieved here by either: (i) the tree level barrier $\mu_{\phi}^{2}$, (ii) cancellation between the thermal mass term $c_{\phi}\phi^{2}T^{2}$ arising once thermal corrections are included in the potential and the negative $\lambda_{ \phi }\phi^{4}$ term. In order for the particle-antiparticle asymmetry to survive we require $\phi_{n}/T_{n} >  g_{G}(1.5 - 1.8)$, where $g_{G}$ is the SU(2)$_{G}$ gauge coupling, $\phi_{n}$ is the field value at the global minimum of the potential at the nucleation temperature $T_{n}$, and the uncertainty comes from the difficulty in determine the prefactor of the sphaleron rate. Using standard techniques one is able to numerically calculate the profile of the critical bubble and also the nucleation temperature $T_{n}$. Bubbles will nucleate when $S_{3}/T \approx 140$, where $S_{3}$ is the O(3) symmetric Euclidean action. An example of the bubble profile is shown in Fig.~\ref{fig:bubble}. Also shown is the evolution of the effective potential with temperature. 

\begin{figure}[t]
	\begin{center}
	\includegraphics[width=165pt]{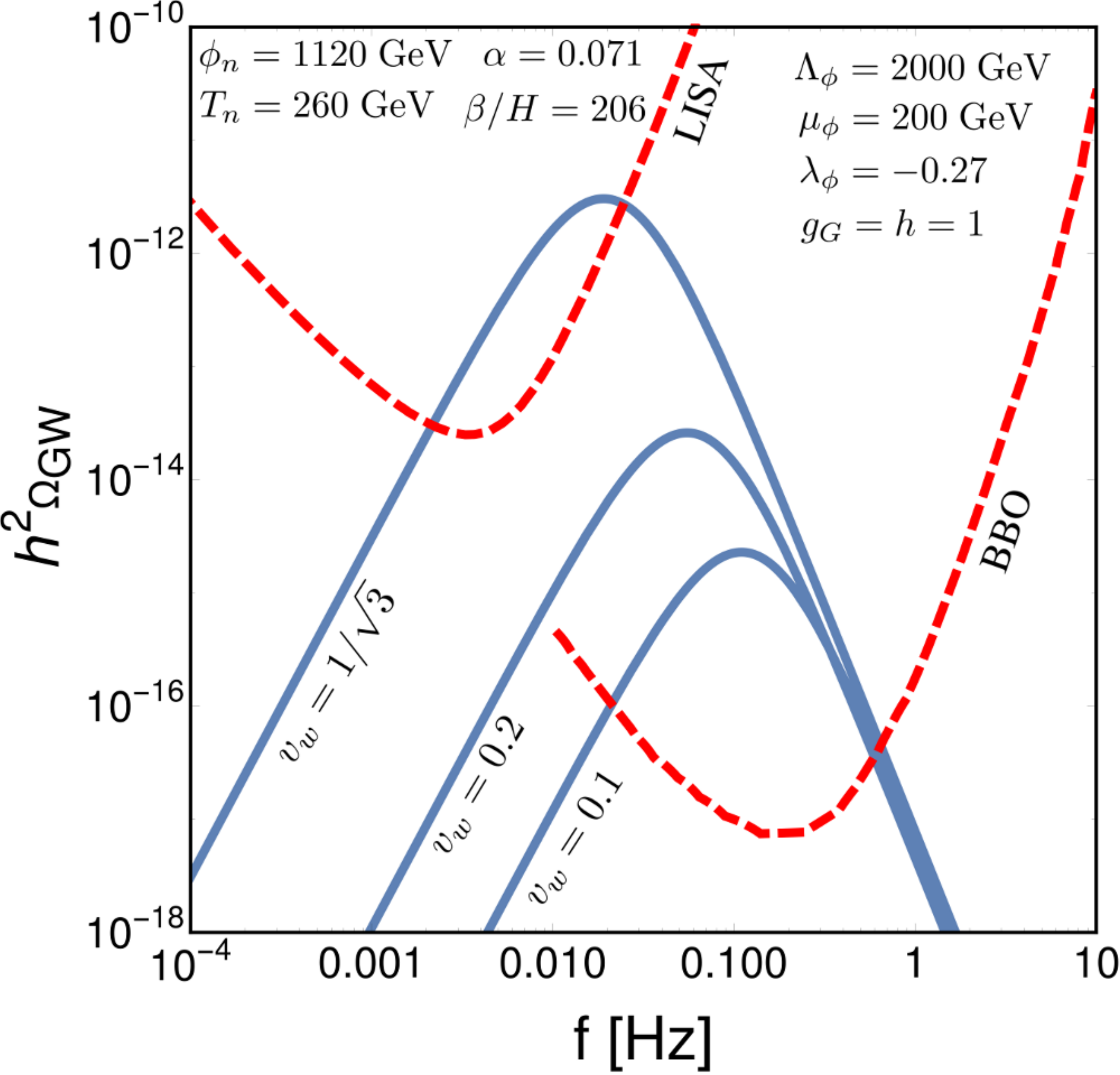} $\qquad$
	\includegraphics[width=155pt]{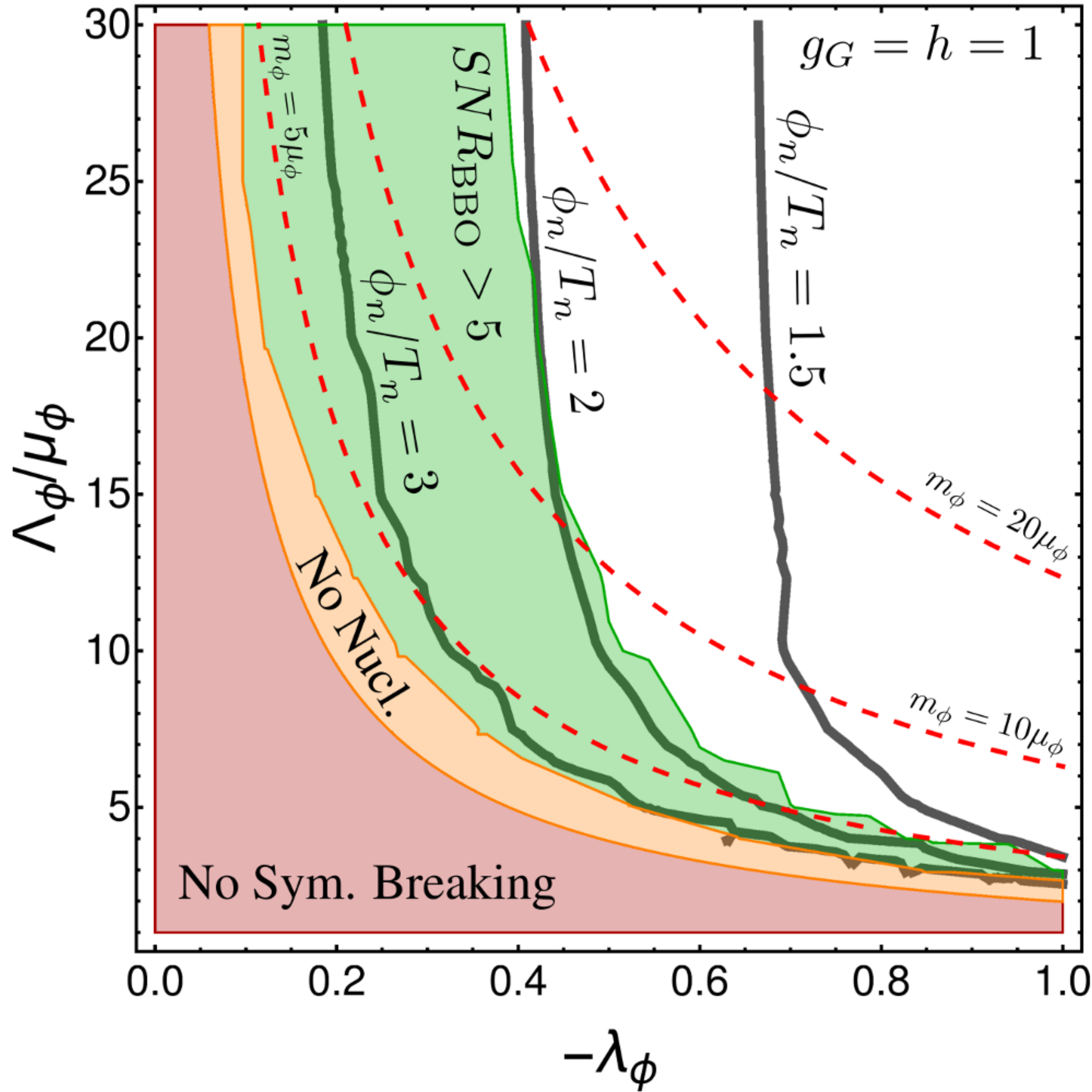}
	\end{center}
\caption{\small Left: An example of the spectrum of graviational waves in this scenario. Right: A scan over the parameter space of the model. Figures from: I. Baldes, {\em JCAP} {\bf 1705}, 28 (2017), {\scriptsize \copyright} SISSA Medialab Srl.. Reproduced by permission of IOP Publishing. All rights reserved.}
\label{fig:spectrum}
\end{figure}

The  sound waves in the plasma are expected to be the dominant contribution to the gravitational wave spectrum, $\Omega_{\rm GW}(f)$, in this scenario.~\cite{Soundwaves} Standard parametrisations taken from simulations are used which allows one to find $\Omega_{\rm GW}(f)$ given $T_{n}$, $\alpha$ (related to the energy released), $\beta/H$ (related to the speed of the phase transition), and the wall velocity $v_{w}$. $\alpha$ and $\beta$ are extracted from our calculation of the critical bubble. While $v_{w}$ is a difficult to determine quantity~\cite{Moore} and has not yet been calculated for this scenario. However, we do  check the Bodeker-Moore criterion~\cite{Bodeker} for a non-runaway wall is fufilled, as runaway walls are inconsistent with this baryogenesis mechanism. Subsonic walls with $v_{w} < 1/\sqrt{3}$ are required to allow the fermions carrying the chiral asymmetry to diffuse back into the symmetric phase where the sphalerons are active. Using $v_{w} = 1/\sqrt{3}$ therefore gives us the most optimistic scenario for a signal. The strong dependence of the gravitational wave signal on the wall velocity is illustrated in Fig.~\ref{fig:spectrum} together with projected sensitivities of LISA~\cite{LISAsens} and BBO~\cite{BBOsens} to stochastic gravitational wave backgrounds. A slice of parameter space showing areas to which BBO is sensitive is also shown.

\section{Other Tests} 
Efficient annihilation of the symmetric DM component is achieved through annihilation to a dark gauge boson. If this dark mediator is massless, stringent, yet somewhat contentious constraints can be derived from Halo ellipticity,~\cite{Halo} which effectively rules out this possibility. The limit from the Planck determination of the effective number of neutrinos, $N_{\rm eff}=3.15 \pm 0.23$, is more robust;~\cite{Planck} by taking into account dark and visible sector degrees-of-freedom to determine the dark photon temperature, one finds the model is still marginally consistent at $\sim 2\sigma$. If the dark photon gains a mass, $M_{D}$, either through the St\"{u}ckelberg or Higgs mechanism, the Halo ellipticity and $N_{\rm eff}$ constraints can be avoided. In this case one finds a direct detection cross section 
	\begin{equation}
	\sigma_{D}^{\rm SI} \sim 10^{-40} \; \mathrm{cm}^{2} \; \left( \frac{\epsilon}{10^{-5}} \right)^{2} \left( \frac{\alpha_{D}}{10^{-2}} \right) \left( \frac{300 \; \mathrm{MeV} }{ M_{D} } \right)^{4}\left( \frac{\mu_{N}}{0.6 \; \mathrm{GeV}} \right)^{2},
	\end{equation}
where $\epsilon$ is the kinetic mixing parameter, $\alpha_{D}$ is the dark fine structure constant and $\mu_{N}$ is the DM-Nucleon reduced mass. For $M_{DM}=1.5$ GeV, CRESST-II sets a bound $\sigma_{D}^{\rm SI}  < 2.7 \times 10^{−39} \; \mathrm{cm}^{2}$ while the neutrino floor is at $\approx 10^{-43} \; \mathrm{cm}^{2}$, allowing further parameter space to be probed.~\cite{Cresst}

\section{Conclusions}
We have investigated the phenomenology of an asymmetric dark matter scenario in which the visible and DM densities are a consequence of electroweak-style genesis in an exotic phase transition. The resulting stochastic gravitational wave background offers a signal of the high scale generative sector while direct detection offers a possibility for probing the current day dark sector.

\section*{Acknowledgments}
I thank the organisers and in particular, D. Samtleben, for the invitation to speak at the conference and the conference participants for the engaging atmosphere.


\section*{References}

\begin{thebibliography}{99}

\bibitem{ADMreview}K. Petraki and R.R. Volkas, \Journal{{\em Int. J. Mod. Phys.} A}{28}{1330028}{2013}.

\bibitem{Kallia}K. Petraki, M. Trodden and R.R. Volkas, \Journal{{\em JCAP}}{1202}{44}{2012}.

\bibitem{Soundwaves}M. Hindmarsh {\it et al}, \Journal{\PRL}{116}{231101}{2016}.

\bibitem{Moore}G.D. Moore and T. Prokopec, \Journal{\PRD}{52}{7182}{1995}.

\bibitem{Bodeker}D. Bodeker and G.D. Moore, \Journal{{\em JCAP}}{0905}{009}{2009}.

\bibitem{LISAsens}C. Caprini {\it et al}, \Journal{{\em JCAP}}{1604}{001}{2016}.

\bibitem{BBOsens}E. Thrane and J. D. Romano, \Journal{\PRD}{88}{124032}{2013}.

\bibitem{Halo}K. Petraki, L. Pearce and A. Kusenko, \Journal{{\em JCAP}}{1407}{039}{2014}.

\bibitem{Planck}P. A. R. Ade {\it et al}, \Journal{{\em Astron. Astrophys.} A}{13}{594}{2016}.

\bibitem{Cresst}G. Angloher {\it et al}, \Journal{{\em Eur. Phys. J.} C}{76}{25}{2016}.

\end{thebibliography}
\end{document}